\documentclass[twocolumn,showpacs,preprintnumbers,amsmath,amssymb]{revtex4}
\usepackage{dcolumn}
\usepackage{bm}

\newcommand{\ket}[1]{\displaystyle{|#1\rangle}}
\newcommand{\bra}[1]{\displaystyle{\langle #1|}}
\newcommand{\Er}{\mathbf{E}(\mathbf{r})}

\newcommand{\skj}{\sum_{\mathbf{k}j}}
\newcommand{\skkjj}{\sum_{\mathbf{k}\mathbf{k}'jj'}}
\newcommand{\skkjjppll}{\sum_{\substack{\mathbf{k}\mathbf{k}'jj'\\\mathbf{p}\mathbf{p}'ll'}}}
\newcommand{\Ekj}{\mathbf{E}_{\mathbf{k}j}}
\newcommand{\Ekjp}{\mathbf{E}_{\mathbf{k}'j'}}
\newcommand{\akj}{a_{\mathbf{k}j}}
\newcommand{\ackj}{a_{\mathbf{k}j}^\dag}
\newcommand{\akjp}{a_{\mathbf{k}'j'}}
\newcommand{\ackjp}{a_{\mathbf{k}'j'}^\dag}
\newcommand{\apil}{a_{\mathbf{p}l}}
\newcommand{\acpl}{a_{\mathbf{p}l}^\dag}
\newcommand{\aplp}{a_{\mathbf{p}'l'}}
\newcommand{\acplp}{a_{\mathbf{p}'l'}^\dag}
\newcommand{\fkjr}{\mathbf{f}(\mathbf{k}j,\mathbf{r})}
\newcommand{\ekj}{\mathbf{e}_{\mathbf{k}j}}

\begin{document}
\preprint{APS/123-QED}
\title{Fluctuations of the Casimir-Polder force between an atom and a conducting wall}
\author{R. Messina and R. Passante}
\email{roberto.passante@fisica.unipa.it}\affiliation{Dipartimento di
Scienze Fisiche ed Astronomiche dell'Universit\`{a} degli Studi di
Palermo and CNSIM,\\Via Archirafi 36, 90123 Palermo,
Italy}\date{\today}
\begin{abstract}We consider the quantum fluctuations of the Casimir-Polder force
between a neutral atom and a perfectly conducting wall in the
ground state of the system. In order to obtain the atom-wall force
fluctuation we first define an operator directly associated to the
force experienced by the atom considered as a polarizable body in
an electromagnetic field, and we use a time-averaged force
operator in order to avoid ultraviolet divergences appearing in
the fluctuation of the force. This time-averaged force operator
takes into account that any measurement involves a finite time. We
also calculate the Casimir-Polder force fluctuation for an atom
between two conducting walls. Experimental observability of these
Casimir-Polder force fluctuations is also discussed, as well as
the dependence of the relative force fluctuation on the duration
of the measurement.
\end{abstract}


\maketitle

\section{Introduction}
A striking consequence of quantum electrodynamics is that the
radiation field, even in its ground state, has fluctuations of the
electric and magnetic fields around the zero value
\cite{Milonni,CPP}. This theoretical prediction has many
remarkable observable consequences. One of them is the prediction
of the existence of electromagnetic forces between two or more
uncharged objects in the vacuum. The existence of these forces was
first predicted in two papers by Casimir \cite{Casimir} and by
Casimir and Polder \cite{CasPol} in $1948$, and, from then
onwards, the interest on this subject has grown exponentially.
Many experiments have definitively proved these effects with
remarkable precision, measuring the Casimir force between a lens
and a wall \cite{Lamo,MohRoy}, between a neutral atom and a wall
\cite{Sukenik1,Sukenik2,DruzhDeKie}, between a surface and a
Bose-Einstein condensate \cite{HOMC05,OWAPSC07} and between two
metallic neutral parallel plates \cite{Bressi1,Bressi2}.

One aspect of Casimir-Polder forces has not received, in our
opinion, enough attention: the value calculated for the force is
actually an average value and it may in principle exhibit quantum
fluctuations. The study of fluctuations of Casimir-Polder forces
could be relevant for the stability of MEMS and NEMS, which are
devices based on controlling the movement of metallic objects
separated by distances of the order of micrometers or nanometers,
where Casimir forces may be relevant \cite{MEMS1,MEMS2}.

Casimir and Casimir-Polder force fluctuations have been studied,
with different approaches, by G. Barton
\cite{Barton1,Barton2,Barton94}, Jaekel and Reynaud \cite{JR92},
and C.H. Wu et al \cite{WKF02,WF01}. Our approach follows that of
Barton, with the difference that, whereas Barton studied only
entirely macroscopical systems, we apply his method of
time-averaged operators to the study of systems with also one atom
present.

In this paper we calculate the fluctuations of the Casimir-Polder
force between a neutral atom and a perfectly conducting wall in
the ground state of the system. We first introduce an operator
directly associated with the force experienced by a polarizable
body in an electromagnetic field. Since the quadratic mean value
of the force proves to be divergent, we make use of the method of
time-averaged operators introduced and widely used by Barton in
his papers about fluctuations of Casimir forces for macroscopic
bodies \cite{Barton1,Barton2,Barton94}. The analytical techniques
used are introduced in Section \ref{SezOperatore}, whereas the
detailed calculation is given in Section \ref{SezCalcolo}. In
Section \ref{SezDuePareti} the Casimir-Polder force fluctuation is
obtained in the case of an atom between two parallel walls: this
permits us to specialize our results to a system for which the
atom-wall Casimir-Polder force has been measured with precision
\cite{Sukenik1,Sukenik2}. In the Conclusions, we make further
remarks on our results and outline possible future developments.

\section{The force operator and the method of time-averaged operators}
\label{SezOperatore}Let us first briefly review the method often
used to calculate the average Casimir-Polder force between an atom
and a neutral conducting wall. The calculation is carried out by
considering the interaction energy of the atom with the radiation
field in the vacuum state. A convenient choice is to use the
effective interaction Hamiltonian given by \cite{PassPowThiru}
\begin{equation}
\label{Interaz}
W=-\frac{1}{2}\skkjj\alpha(k)\Ekj(\mathbf{r}_A)\cdot\Ekjp(\mathbf{r}_A)
\end{equation}
where
\begin{equation}\Er=\skj\Ekj(\mathbf{r})=i\skj
\sqrt{\frac{2\pi\hbar\omega_k}{V}}(\akj-\ackj)\fkjr
\end{equation}
$\alpha (k)$ being the dynamical polarizability of the atom and
$\fkjr$ the mode functions used for the quantization of the
electromagnetic field in the presence of the wall. This
Hamiltonian is correct up to order $\alpha\sim e^2$, $e$ being the
electron charge. This effective Hamiltonian allows considerable
simplification in the calculation of Casimir-Polder potentials,
both in stational and dynamical cases \cite{RPP04,SPR06,MP07}. The
presence of the wall is taken into account by considering a
conducting cubic cavity defined by
\begin{equation}
-\frac{L}{2}<x<\frac{L}{2}\quad-\frac{L}{2}<y<\frac{L}{2}\quad
0<z<L
\end{equation}
where $L$ is the side of the cavity and $V=L^3$ its volume. The
mode functions for this box have components \cite{PT82,Milonni}
\begin{widetext}
\begin{equation}
\begin{split}f_x(\mathbf{k}j,\mathbf{r})&=\sqrt{8}(\ekj)_x\cos\Bigl[k_x
\Bigl(x+\frac{L}{2}\Bigr)\Bigr]\sin\Bigl[k_y\Bigl(y+\frac{L}{2}\Bigr)
\Bigr]\sin\bigl(k_z z\bigr)\\
f_y(\mathbf{k}j,\mathbf{r})&=\sqrt{8}(\ekj)_y\sin\Bigl[k_x
\Bigl(x+\frac{L}{2}\Bigr)\Bigr]\cos\Bigl[k_y\Bigl(y+\frac{L}{2}\Bigr)\Bigr]
\sin\bigl(k_z z\bigr)\\
f_z(\mathbf{k}j,\mathbf{r})&=\sqrt{8}(\ekj)_z\sin
\Bigl[k_x\Bigl(x+\frac{L}{2}\Bigr)\Bigr]\sin\Bigl[k_y\Bigl(y+\frac{L}{2}\Bigr)\Bigr]
\cos\bigl(k_z z\bigr)\\\end{split}\end{equation}
\end{widetext}
where $\ekj$ are polarization unit vectors and the allowed values of
$\mathbf{k}$ have components
\begin{equation}
k_x=\frac{l\pi}L,\quad k_y=\frac{m\pi}L, \quad
k_z=\frac{n\pi}L,\quad l,m,n=0,1,2,\dots
\end{equation}
We obtain a correct description of a conducting wall located in
$z=0$ by taking the limit $L \to +\infty$.

We now calculate the quantum average of the operator
\eqref{Interaz} on the ground state $\ket{0}$ of the
electromagnetic field. If we consider the atom located in
$\mathbf{r}_A=(0,0,d)$, with $d>0$, we obtain
\begin{equation}\begin{split}E(d)&=\bra{0}W\ket{0} \nonumber\\
&=-\frac{\pi\hbar c}{V}\skj
k\alpha(k)\Bigl[\mathbf{f}(\mathbf{k}j,\mathbf{r}_A)
\cdot\mathbf{f}(\mathbf{k}'j',\mathbf{r}_A)\Bigr].\\\end{split}\end{equation}
Because this interaction energy depends on the $z$-coordinate of the
atom, in a quasi-stationary approach the atom experiences a force
given by
\begin{equation}F_A(d)=-\frac{\partial}{\partial d}E(d).\end{equation}

Using the explicit expression of the mode functions $\fkjr$ it is
easy to get the result
\begin{equation}\label{ValoreForza}F_A(d)=-\frac{3\hbar c\alpha}{2\pi d^5}\end{equation}
where $d$ is the atom-wall distance, $\alpha$ is the static
polarizability of the atom and the minus sign indicates that the
force is attractive. The expression \eqref{ValoreForza} is valid in
the so-called \emph{far zone} defined by the condition
$d>>c/\omega_0$, $\omega_0$ being a typical atomic transition
frequency. This result coincides with that obtained by Casimir and
Polder \cite{CasPol}. Effects related to a possible motion of the
atom have been recently considered in the literature by inclusion of
the atomic translational degrees of freedom \cite{SHP03,HRS04}.

This method provides a physically transparent way for calculating
the average force on the atom but it does not enable to easily
obtain the quadratic average value of the force, necessary for the
fluctuation. Thus we introduce a new operator associated to the
force on the atom. In order to define such an operator we formally
take minus the derivative of the operator \eqref{Interaz} with
respect to the $z$-coordinate of the atom $d$, treated as a
parameter. So we take the following quantity as the force operator
\begin{widetext}
\begin{equation}\label{OpForza}F_A=-\frac{\partial}{\partial d}W=-\frac{\pi\hbar
c}{V}\skkjj\sqrt{kk'}\alpha(k)(\akj-\ackj)(\akjp-\ackjp)
F_A(\mathbf{k}j,\mathbf{k}'j',d)\end{equation}
where
\begin{equation}F_A(\mathbf{k}j,\mathbf{k}'j',d)=
\frac{\partial}{\partial d}
\Bigl[\mathbf{f}(\mathbf{k}j,\mathbf{r}_A)
\cdot\mathbf{f}(\mathbf{k}'j',\mathbf{r}_A)\Bigr].\end{equation}
\end{widetext}

It is easy to show that the quantum average of the operator
\eqref{OpForza} on the vacuum state $\ket{0}$ gives back the
expression \eqref{ValoreForza} of the force, since the derivation
with respect to $d$ commutes with the quantum average. This force
operator is correct up to order $\alpha$. We can now consider the
operator corresponding to the square of the force, that is
\begin{widetext}\begin{equation}\begin{split}F_A^2=
\Bigl(\frac{\pi\hbar c}{V}\Bigr)^2\skkjjppll&\sqrt{kk'pp'}
\alpha(k)\alpha(p)(\akj-\ackj)(\akjp-\ackjp)\\
&\times(\apil-\acpl)(\aplp-\acplp)
F_A(\mathbf{k}j,\mathbf{k}'j',d)F_A(\mathbf{p}l,\mathbf{p}'l',d).
\\\end{split}\end{equation}\end{widetext}
Using this operator, however, we find that the average squared
value of the force has an ultraviolet divergence that cannot be
regularized by a cut-off function. An analogous problem was
encountered by Barton in his works on force fluctuations for
macroscopic bodies \cite{Barton1,Barton2,Barton94}. In order to
solve this problem, he proposed to consider explicitly that every
real measurement must involve a finite time $T$ and thus
considered a temporal average of the force. The basic idea is to
integrate on time the quantum average value with a response
function $f(t)$ describing the measurement process. Then, being
$F_A$ the force operator in the Schr\"{o}dinger representation and
$H$ the Hamiltonian of the system, the time-averaged force with an
instrument characterized by a normalized response function $f(t)$
is given by
\begin{equation}
\label{forzamedia}\begin{split}\overline{F_A}(d)&=
\int_{-\infty}^{+\infty}dt\,f(t)\bra{0}F(t)\ket{0} \\
&=\int_{-\infty}^{+\infty}dt\,f(t)\bra{0}e^{\frac{i}{\hbar}Ht}
Fe^{-\frac{i}{\hbar}Ht}\ket{0}.\\\end{split}\end{equation}

The expression \eqref{forzamedia}  can be thought as the quantum
average on the state $\ket{0}$ of the \emph{time-averaged
operator}
\begin{equation}\label{FMediaDef}\overline{F_A}=
\int_{-\infty}^{+\infty}dt\,f(t)e^{\frac{i}{\hbar}Ht}
F_Ae^{-\frac{i}{\hbar}Ht}\end{equation}
which is a time independent operator whose definition depends on the
properties of the instrument used for the measurement.

The choice of using the operator $\overline{F_A}$ does not change
our results for the average force, whereas it introduces, as we
will show in the next Section, a natural frequency cutoff in the
average squared value of the force, such as $e^{-\omega T}$. This
is indeed reasonable since an instrument with an integration time
$T$ does not resolve processes with frequencies larger than
$T^{-1}$.

\section{The fluctuation of the Casimir-Polder force between an atom and a wall}
\label{SezCalcolo}We now use eq.\eqref{FMediaDef} to define the
time-averaged operator associated to the square of the force, which
is
\begin{widetext}
\begin{equation}
\label{FQuadraMediaDef}\Bigl(\overline{F_A}\Bigr)^2
=\int_{-\infty}^{+\infty}dt\,f(t)\int_{-\infty}^{+\infty}dt'\,f(t')
e^{\frac{i}{\hbar}Ht}F_Ae^{-\frac{i}{\hbar}H(t-t')}F_Ae^{-\frac{i}{\hbar}Ht'}.
\end{equation}
\end{widetext}
In this equation, $H$ is the total Hamiltonian of the system, i.e.
$H=H_F+H_A+W$, where $H_F$ and $H_A$ are, respectively, the
Hamiltonian of the free electromagnetic field and of the atom, and
$W$ is the interaction term introduced in the previous Section. We
have obtained, for the mean force, a result correct to the first
order in the polarizability $\alpha$ of the atom. As a
consequence, a coherent result for the average value of the square
of the force should contain the second power of $\alpha$. Since
$F_A$ is an operator of order $\alpha$, it is clear from
\eqref{FQuadraMediaDef} that we must retain only $H_F+H_A$ instead
of $H$ in the exponentials, in order to have a mean quadratic
value of $\overline{F_A}$ proportional to $\alpha^2$. Besides, as
the state $\ket{0}$ does not contain atomic variables, it is
sufficient to put $H=H_F$ in \eqref{FQuadraMediaDef}.

Thus, taking the response function $f(t)$ as a lorentzian of width
$T$, we obtain the following expression for the fluctuation
$\Delta \overline{F_A}= (\langle F_A^2 \rangle - \langle F_A
\rangle^2)^{1/2}$ of the Casimir-Polder force on the atom in far
zone
\begin{widetext}
\begin{equation}\begin{split}\Delta
\overline{F_A}=\frac{\hbar
c\alpha}{4\pi}\,\frac1{c^5T^5\left(1+\left(\frac{cT}d\right)^2\right)^4}\,&\Biggl(
5+40\left(\frac{cT}d\right)^2+
145\left(\frac{cT}d\right)^4+317\left(\frac{cT}d\right)^6 \\
&\,+400\left(\frac{cT}d\right)^8+285\left(\frac{cT}d\right)^{10}+10\left(\frac{cT}d\right)^{12}+
86\left(\frac{cT}d\right)^{14}\Biggr)^{1/2}\\\end{split}\end{equation}\end{widetext}
We can easily study the behavior of the relative fluctuation, that
is the standard deviation of the force divided by the absolute
value of the average force, in two different limiting cases. When
$d<<cT$ we get
\begin{equation}\frac{\Delta\overline{F_A}}{|\bra{0}F_A\ket{0}|}=
\frac{1}{3}\sqrt{\frac{43}{2}}\Bigl(\frac{d}{cT}\Bigr)^6\end{equation}
whereas in the case $d>>cT$ we have
\begin{equation}\frac{\Delta\overline{F_A}}{|\bra{0}F_A\ket{0}|}=
\frac{\sqrt{5}}{6}\Bigl(\frac{d}{cT}\Bigr)^5.\end{equation} In the
first case, the force fluctuation seems to be negligible compared
to the average force, whilst in the second case it would result
much larger than the average force, thus experimentally
observable.

When the atom-wall distance is of the order of $d \sim 1\,\mu m$
(typical distance in actual experimental setups \cite{Lamo}) the
timescale which separates the two regimes is $T \sim 10^{-14}\,s$,
which is a very short timescale but probably no longer impossible
nowadays. Therefore, to evaluate the experimental observability of
the fluctuations we need a reasonable value of the measurement
time $T$. In order to compare our theoretical predictions for the
force fluctuations with actual precision measurements of the
atom-wall Casimir-Polder force in the far zone, we have extended
our calculations to the system of one atom between two parallel
metallic walls. In fact, for this system well-established
precision measuements exist \cite{Sukenik2}.

\section{Fluctuations of the Casimir-Polder on an atom between two conducting walls}
\label{SezDuePareti}
In order to take into account the presence of
the two parallel walls separated by a distance $L$, we make use of
the mode functions associated to a conducting parallelepiped
cavity defined by
\begin{equation}
-\frac{L_1}{2}<x<\frac{L_1}{2}\quad-\frac{L_1}{2}<y<\frac{L_1}{2}\quad
-\frac L2<z<\frac L2\end{equation} which are easily found to be
\begin{widetext}
\begin{equation}
\begin{split}f_x(\mathbf{k}j,\mathbf{r})&=\sqrt{8}(\ekj)_x\cos\Bigl[k_x
\Bigl(x+\frac{L_1}{2}\Bigr)\Bigr]\sin\Bigl[k_y\Bigl(y+\frac{L_1}{2}\Bigr)
\Bigr]\sin\Bigl[k_z\Bigl(z+\frac{L}{2}\Bigr)\Bigr]\\
f_y(\mathbf{k}j,\mathbf{r})&=\sqrt{8}(\ekj)_y\sin\Bigl[k_x
\Bigl(x+\frac{L_1}{2}\Bigr)\Bigr]\cos\Bigl[k_y\Bigl(y+\frac{L_1}{2}\Bigr)\Bigr]
\sin\Bigl[k_z\Bigl(z+\frac{L}{2}\Bigr)\Bigr]\\
f_z(\mathbf{k}j,\mathbf{r})&=\sqrt{8}(\ekj)_z\sin
\Bigl[k_x\Bigl(x+\frac{L_1}{2}\Bigr)\Bigr]\sin\Bigl[k_y\Bigl(y+\frac{L_1}{2}\Bigr)\Bigr]
\cos\Bigl[k_z\Bigl(z+\frac{L}{2}\Bigr)\Bigr].\\\end{split}\end{equation}
\end{widetext}
In the limit $L_1\to+\infty$ we obtain two infinite conducting
walls located in $z=\pm L/2$. Following the same steps of Sections
\ref{SezOperatore} and \ref{SezCalcolo} we obtain the following
expression for the average force on the atom
\begin{equation}F_A(d)=-\frac{\pi^4\hbar c\alpha}{8L^5}\,\frac{\sin\bigl(\frac{3\pi
d}{L}\bigr)-11\sin\bigl(\frac{\pi
d}{L}\bigr)}{\cos^5\bigl(\frac{\pi d}{L}\bigr)}\end{equation}
where $L$ is the distance between the two walls and $-L/2<d< L/2$
is the distance of the atom from the plane in the middle of the
plates. This force vanishes for $d=0$ for symmetry reasons. This
result coincides with a result already obtained by Barton
\cite{BartonDuePareti}. We have then calculated, using the
time-averaged operator method described in Section
\ref{SezOperatore}, the value of the relative fluctuation of the
force. We find that also in this case the relative fluctuation
depends on the measurement time. Since the experiment in
\cite{Sukenik2} consists in the passage of a beam of atoms between
the two walls, an estimate of the integration time $T$ can be
obtained from the length of the cavity ($8\,\mbox{mm}$ in the
mentioned experiment) and the average speed of the atoms. This
average velocity can be easily obtained from the Maxwell-Boltzmann
distribution of the atoms, and thus we get
$T\sim10^{-5}\,\mbox{s}$. In this case the expression of the
relative fluctuation can be simplified, yielding
\begin{widetext}\begin{equation}\frac{\Delta
F_A}{|\bra{0}F_A\ket{0}|}\simeq \frac{e^{-\pi cT/L}}{\left(\frac
{2\pi cT}L \right)^{5/2}}\frac{\cos^6(10^6\pi
d\,\mbox{m}^{-1})}{\Bigl|\sin(3\cdot10^6\pi
d\,\mbox{m}^{-1})-11\sin(10^6\pi
d\,\mbox{m}^{-1})\Bigr|}\end{equation}\end{widetext} where we have
used the fact that $L=1 \, \mbox{$\mu m$}$ and units for the
distance $d$ have been excplicitly specified. This function
diverges for $d\to0$ (since the average force vanishes for $d=0$),
but is already negligible for $d\sim10^{-10}\,\mbox{m}$, that is
at a distance of the order of the Bohr radius. Consequently, we
can conclude that in this experimental setup the fluctuation of
the force is so small to be hardly observable. This does not
exclude observability of the fluctuation of the Casimir-Polder
force in future experimental setups characterized by shorter
measurement times, of course.

\section{Conclusions}
In this paper we have considered the fluctuation of the
Casimir-Polder force experienced by a neutral atom in front of an
uncharged conducting wall or between two parallel uncharged walls.
We have first introduced a quantum operator directly associated to
the force on the atom, considered as a microscopic polarizable
body, due to the electromagnetic field. This operator has been
obtained by taking minus the derivative of the operator
corresponding to the atom-field effective interaction energy with
respect to the coordinate of the atom normal to the plate(s). This
operator has been used to calculate the mean force in both
configurations. As for the quadratic mean value, in order to go
beyond the non-regularizable ultraviolet divergences encountered,
we have used the method of time-averaged operators, previously
used by Barton for the Casimir force fluctuation between
macroscopic bodies. We have obtained the relative fluctuation both
in the cases of one and two walls. In the case of one wall, the
value of the relative force fluctuation strongly depends on the
ratio between the atom-wall distance $d$ and the distance $cT$
travelled by the light during the measurement time $T$.
Fluctuations are larger the smaller is the duration of the force
measurement. In the case of two walls, we have been also able to
estimate the experimental observability of this fluctuation in a
recent precision experiment on the atom-wall Casimir-Polder force
in the far zone \cite{Sukenik2}, concluding that in this
experiment the fluctuations are very small and hardly observable.
Our results show that force fluctuations should however be
observable in experiments in which the force is measured in much
shorter timescales. Future extensions of this work involve the
calculation of the Casimir-Polder force between two atoms
(retarded van der Waals force), where one may expect that the
relative fluctuation of the force could be significantly larger
because only microscopic objects are involved.

\begin{acknowledgments}
Partial support by Ministero dell'Universit\`{a} e della Ricerca
Scientifica e Tecnologica and by Comitato Regionale di Ricerche
Nucleari e di Struttura della Materia is also acknowledged.
\end{acknowledgments}

\end{document}